\documentclass[nonacm,sigconf]{acmart}
\usepackage{amsmath,amsfonts} 
\usepackage{algorithmic}
\usepackage{graphicx}
\usepackage{textcomp}
\usepackage{xcolor}
\usepackage[utf8]{inputenc}
\usepackage[english]{babel}
\usepackage{babel}
\usepackage{enumitem}
\usepackage{titlesec}
\usepackage{tabularx}
\usepackage[switch,modulo]{lineno}

\titleclass{\subsubsubsection}{straight}[\subsection]
\newcounter{subsubsubsection}[subsubsection]
\renewcommand\thesubsubsubsection{\thesubsubsection.\arabic{subsubsubsection}}
\titleformat{\subsubsubsection}[runin]{\normalfont\normalsize\bfseries}{\thesubsubsubsection}{1em}{}[.]
\titlespacing*{\subsubsubsection}{0pt}{3.25ex plus 1ex minus .2ex}{1.5ex plus .2ex}

\usepackage{soul}

\begin{document}

\title{Towards Deep Learning Enabled Cybersecurity Risk Assessment for Microservice Architectures}

\author{Majid Abdulsatar}
\affiliation{%
  \institution{School of Computer and Mathematical Sciences}
  \country{University of Adelaide, Australia}
  }
\email{majid.abdulsatar@student.adelaide.edu.au}

\author{Hussain Ahmad$^{\ast}$}
\affiliation{%
  \institution{School of Computer and Mathematical Sciences}
  \country{University of Adelaide, Australia}
  }
\email{hussain.ahmad@adelaide.edu.au}
\thanks{*Corresponding author}

\author{Diksha Goel}
\affiliation{%
  \institution{CSIRO's Data61, Australia}
  \country{}
  }
\email{diksha.goel@data61.csiro.au}

\author{Faheem Ullah}
\affiliation{%
  \institution{School of Computer and Mathematical Sciences}
  \country{University of Adelaide, Australia}
  }
\email{faheem.ullah@adelaide.edu.au}

\begin{abstract}

The widespread adoption of microservice architectures has given rise to a new set of software security challenges. These challenges stem from the unique features inherent in microservices. It is important to systematically assess and address software security challenges such as software security risk assessment. However, existing approaches prove inefficient in accurately evaluating the security risks associated with microservice architectures. To address this issue, we propose \textit{CyberWise Predictor}, a framework designed for predicting and assessing security risks associated with microservice architectures. Our framework employs deep learning-based natural language processing models to analyze vulnerability descriptions for predicting vulnerability metrics to assess security risks. Our experimental evaluation shows the effectiveness of \textit{CyberWise Predictor}, achieving an average accuracy of \textbf{92\%} in automatically predicting vulnerability metrics for new vulnerabilities. Our framework and findings serve as a guide for software developers to identify and mitigate security risks in microservice architectures.

\end{abstract}

\keywords{Cybersecurity, Microservices, Deep Learning, Transformers, NLP, Risk Assessment, Risk Prediction}

\maketitle
\pagestyle{plain}

\section{Introduction}

In recent years, there has been widespread adoption of microservice architectures \cite{pimentel2021self}. This is evident from the transition of several notable enterprises (e.g., Amazon and Netflix) and military systems (e.g., C4ISR systems) from monolithic structures to microservices, all aimed at enhancing the quality of their services \cite{rossi2022dynamic, ahmad2023review}. Microservice architecture is a software architectural style composed of loosely coupled services, each operating independently and focused on a specific role within an application. These services communicate using well-defined Application Programming Interfaces (APIs) \cite{pimentel2021self}. Microservice architectures offer several advantages over traditional monolithic architectures, including autonomous scalability, agility, and reusability \cite {ahmad2024smart}. The independent scalability enables scaling of each service rather than the entire application, whereas the agility of microservice architecture allows each service to operate on its individual development and deployment cycle, facilitating a swift and efficient release of new features. The loosely coupled feature of microservices enhances application resilience against failures and simplifies management compared to traditional monolithic counterparts \cite{blinowski2022monolithic, bushong2021microservice}.

Although microservices offer notable advantages, they also present cybersecurity challenges with their intricate architecture, inter-service communication issues, and dynamic nature \cite{mateus2021security}. Additionally, the integration of third-party components adds complexity, and as the number of services in an application rises, it expands the potential attack surface for malicious actors \cite{dragoni2017microservices, goel2018overview}. Several cybersecurity concerns specific to microservices are reported in \cite{chandramouli2019microservices}, and early research indicates that applying standard patterns for system reliability must now consider new parameters, such as the locations where these patterns are deployed \cite{montesi2018decorator}. With the burgeoning shift towards microservice architectures, a comprehensive analysis of the related cybersecurity risks becomes imperative. This is essential for the timely detection, prioritization, and assessment of cybersecurity risks \cite{bushong2021microservice, he2017authentication, RN11}. The challenge lies in unavailable vulnerability assessment metrics, such as the Common Vulnerability Scoring System (CVSS) severity scores \cite{CVSS_Score}, in vulnerability databases, such as the National Vulnerability Database (NVD) \cite{NVD_Database}. These metrics are crucial for conducting an assessment of cybersecurity risks. While previous studies have explored predictive modeling using Machine Learning (ML) for predicting security vulnerability data \cite{RN3}, various constraints can limit their effectiveness. For instance, ML techniques, such as those utilizing Term Frequency-Inverse Document Frequency (TF-IDF) for text vectorization \cite{abubakar2022sentiment}, struggle with capturing long-range dependencies, potentially leading to a lack of overall contextual understanding. For instance, sentences like "The cat sat on the mat" and "The mat sat on the cat" would have the same TF-IDF vector despite conveying different meanings.

To address these concerns, \textit{we introduce \textit{CyberWise Predictor}, a framework for cybersecurity risk prediction and assessment for microservice architectures}. The framework detects cybersecurity vulnerabilities within microservice architectures by employing Vulnerability Assessment and Penetration Testing (VAPT) tools \cite{vegesna2023utilising}. Subsequently, we systematically map the detected vulnerabilities to the NVD database. To address the issue of incomplete security vulnerability data and recognize the importance of preserving context, particularly in predicting vulnerability impacts, we leverage Deep Learning (DL) based Natural Language Processing (NLP) models, specifically transformers \cite{wolf2020transformers, wolf2019huggingface}. Transformers, known for comprehending long-range dependencies from extensive text corpora, have shown effectiveness across various NLP tasks \cite{wolf2020transformers}, making them an optimal choice for our research. Transformer models (e.g., ERT, RoBERTa, and XLNet) utilize word embedding enhanced with attention mechanisms, capturing semantic information and facilitating understanding of word relationships. The fine-tuned transformer model enhances the completeness of risk assessment frameworks by predicting missing data from vulnerabilities' descriptions. In addition, we develop a taxonomy of cybersecurity vulnerabilities specific to microservice architectures. This taxonomy aids in comprehending the origins of cybersecurity risks, recognizing that vulnerabilities can stem from various sources. These sources encompass design errors, misconfigurations, or common defects like bugs, all of which present potential security risks. The main contributions of this paper are summarized as follows:

\begin{itemize}
\item We introduce a taxonomy of cybersecurity vulnerabilities stemming from the unique characteristics of microservice architectures. This taxonomy aids in understanding the origins and development of security vulnerabilities within microservice architectures.

\item We propose \textit{CyberWise Predictor}, a framework designed for the prediction and assessment of cybersecurity risks in microservices architectures. 
\item We evaluate the effectiveness of the \textit{CyberWise Predictor} through a real-world microservice benchmark system, achieving a 92\% accuracy rate in predicting security risks with only 25\% of the available ground truth data.\\
\end{itemize}

\vspace{-0.1in}

We provide the replication package for \textit{CyberWise Predictor} \cite{selfadaptive, finetuner}, comprising all scripts and data required for reproducing, validating, and extending the results discussed in the paper.
\vspace{5pt}

\textbf{Paper Structure.} The structure of the paper is as follows: Section \ref{Section 2} reviews related work, while Section \ref{Section 3} introduces our taxonomy for microservice vulnerabilities. Section \ref{Section 4} presents a detailed discussion of our framework designed for assessing and predicting cybersecurity risks, along with an in-depth exploration of our experiments. The results and analysis of these experiments are provided in Section \ref{Section 5}. Section \ref{Section 6} presents the implications and limitations of our findings. Finally, Section \ref{Section 7} concludes the paper.

\section{Related Work} \label{Section 2}

This section provides an overview of the current state of research in automated risk detection, prediction, and assessment. While several studies have investigated automated risk management techniques, only a few have specifically addressed automated risk management in the absence of vulnerability metrics. Moreover, a significant gap exists in addressing the unique challenges posed by microservice architectures. Many existing approaches either fail to account for the specific considerations of microservices or utilize methodologies that may not be well-suited for microservice architectures.

Shah et al. \cite{shah2014automated} developed NetNirikshak 1.0, an automated Vulnerability Assessment and Penetration Testing (VAPT) tool for assessing applications and services, identifying vulnerabilities, and generating confidential reports sent via email. Blinowski et al. \cite{blinowski2020cve} employed a vulnerability classification scheme for IoT devices grounded in real-world data. They initially categorized vulnerabilities into seven groups, followed by further classification using standard descriptors from the Common Platform Enumeration (CPE). Leveraging ML techniques, they aimed for automatic classification to address the risks associated with new vulnerabilities. Guo et al. \cite{guo2005automated} proposed a lightweight virtual machine solution for secure vulnerability testing. This approach ensures the safety of testing by creating an exact duplicate of the production-mode network service, while maintaining complete isolation. Moreover, their system can automate the entire vulnerability testing process, allowing for frequent and automatic assessments. Ge et al. \cite{ge2017framework} introduced a paradigm for modeling and evaluating IoT security, which involves developing a graphical security model and a security evaluator to automate security analysis. The security evaluator utilizes various security metrics and outputs analysis results through Symbolic Hierarchical Automated Reliability and Performance Evaluator (SHARPE) \cite{sahner2012performance}, an analytic modeling and assessment tool. However, none of these studies addressed security vulnerability assessment in the absence of vulnerability data.

Duan et al. \cite{RN3} proposed a framework for automating security assessments in IoT networks, utilizing machine learning and natural language processing to analyze vulnerability descriptions and predict metrics such as CVSS score. Singh et al. \cite{RN9} presented a risk estimation model that relies on the NVD and the CVSS. This model, known as the CVSS Risk Level Estimation Model, assesses security risk levels by combining information on vulnerability exploitation duration and occurrence frequency to evaluate the impact derived from CVSS scores. Cam et al. \cite{RN10} introduced a modeling and dynamic analysis approach for real-time assessment of cyber risk within a network. This method represents the network's vulnerabilities, exploitations, and impact dynamically by integrating Markov models and Bayesian networks. Yang et al. \cite{yang2020better} primarily focus on reference information within Common Vulnerabilities and Exposures (CVEs), while Edkrantz et al. \cite{edkrantz2015predicting} incorporate common words and n-grams to enhance classification. Bullough et al. \cite{bullough2017predicting} employ functional engineering techniques and explore the conversion of CVSS v2 to v3 \cite{nowak2021conversion} to expand the vulnerability base. While none of the approaches described above explicitly discusses their application to microservice architectures, we explore security vulnerability assessment within microservice architectures in the absence of vulnerability data. 

Our proposed framework distinguishes itself from existing methodologies in the literature by specifically targeting the lack of vulnerability data within microservice architectures. Our framework utilizes a deep learning-based natural language processing model transformer. Transformers, known for their capacity to comprehend extensive text corpora and grasp long-range dependencies, have demonstrated effectiveness across various natural language processing tasks such as text classification, language translation, and text summarization \cite{wolf2020transformers}. Our model takes CVE descriptions as input to predict CVSS metrics for microservice vulnerabilities. 

\begin{figure}[t]
\centering
\includegraphics[width=1.85\columnwidth]{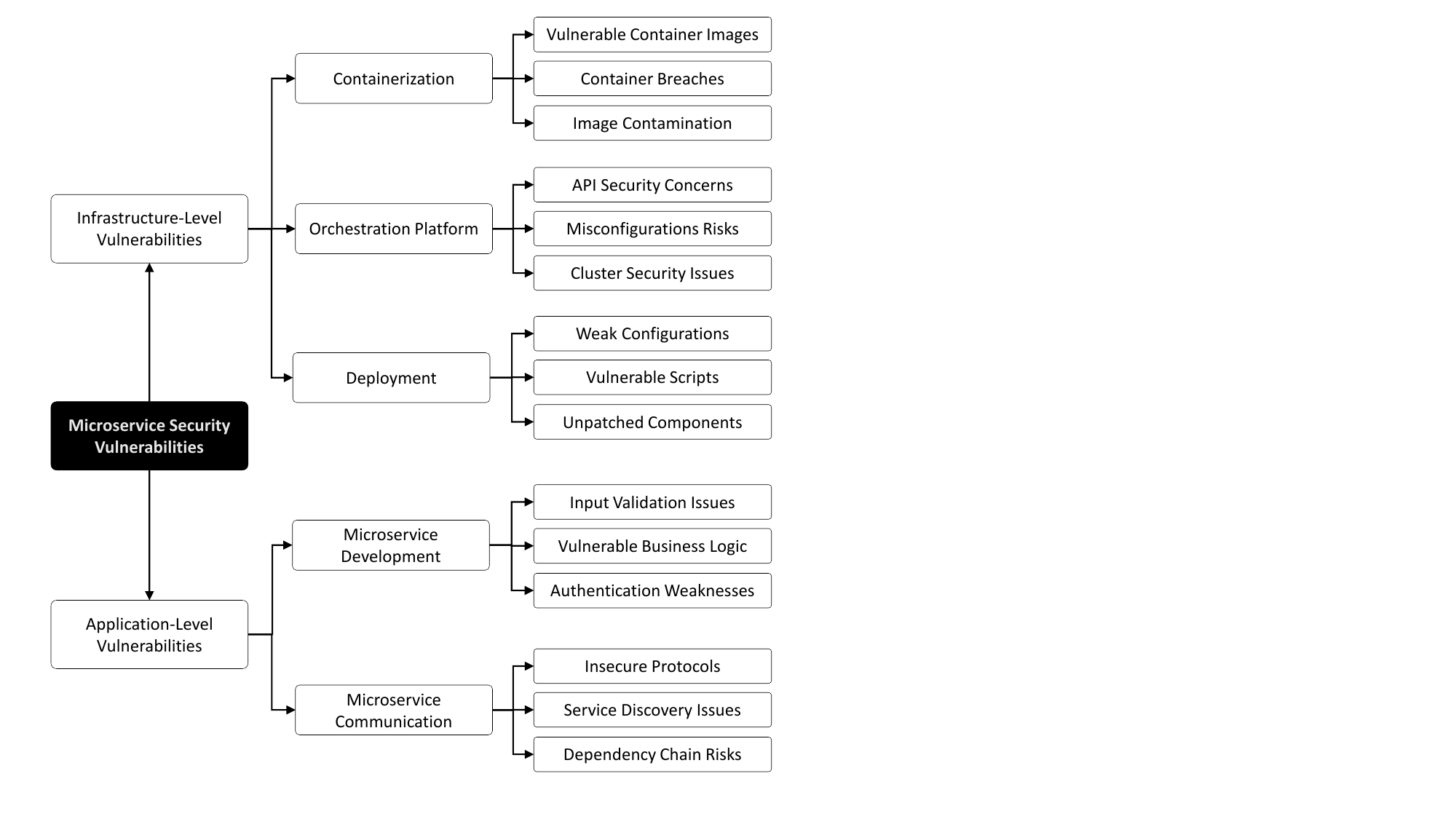}
\caption{Taxonomy of microservice vulnerabilities.}
\label{taxo}
\end{figure}

\section{Microservice Vulnerabilities Taxonomy} \label{Section 3}

To comprehend the roots of security vulnerabilities within microservice architectures, we develop a taxonomy specifically addressing vulnerabilities inherent in microservices. Figure \ref{taxo} presents the hierarchical structure of this taxonomy. The microservice vulnerabilities are classified into two broad categories, with each category further subdivided into more specific types of vulnerabilities.

\subsection{Infrastructure-Level Vulnerabilities}
Infrastructure-related vulnerabilities refer to weaknesses or flaws in the underlying infrastructure components that support the deployment, operation, and communication of microservices \cite{hannousse2021securing}. These vulnerabilities can be exploited by attackers to compromise the confidentiality, availability, or integrity of microservices. \\

\textbf{Container Vulnerabilities.}  Containers encapsulate microservices and their dependencies, providing a lightweight and isolated execution environment. However, vulnerabilities within containers expose microservices, the host system, and the entire infrastructure to various security risks \cite{sultan2019container}. \textit{Vulnerable Container Images.} The utilization of container images containing outdated software or identifiable vulnerabilities poses a risk of exploitation for microservices \cite{ibrahim2019attack}. \textit{Container Breaches.} Inadequate container isolation or misconfigurations can enable attackers to breach containers and gain access to the underlying host system \cite{hannousse2021securing}. \textit{Image Contamination.} Malicious entities might introduce malware or backdoors into container images, thereby jeopardizing the security integrity of microservices.

\textbf{Orchestration Platform Vulnerabilities.} Orchestration platforms, such as Kubernetes, Docker Swarm, or Apache Mesos, are responsible for managing and orchestrating the deployment, scaling, and networking of microservices \cite{ahmad2024smart}. Vulnerabilities within orchestration platforms can be exploited by attackers to compromise the availability, confidentiality, or integrity of the containerized applications or the underlying infrastructure \cite{yarygina2018overcoming}. \textit{API Security Concerns.} Vulnerabilities within the APIs utilized for container management and orchestration, such as the Kubernetes API, have the potential to grant unauthorized access or control over the infrastructure. \textit{Misconfigurations Risks.} Improper configurations of orchestration platforms can lead to insecure deployments, exposing sensitive data, or implementing weak authentication mechanisms \cite{rahman2023security}. \textit{Cluster Security Issues.} Weaknesses in cluster security configurations or the absence of network segmentation may facilitate lateral movement and privilege escalation within the infrastructure \cite{budigiri2021network}.

\textbf{Deployment Vulnerabilities.} Microservice deployment vulnerabilities refer to weaknesses in the process of deploying microservices onto production or operational environments. These vulnerabilities can arise at various stages of the deployment lifecycle and can be exploited by attackers. Given the distributed nature of microservices architectures, deployment vulnerabilities are critical to address as they can have cascading effects on the entire system. \textit{Weak Configuration Management}, including insecure default settings and misconfigurations, exposes vulnerabilities attackers exploit, such as leaving unnecessary ports open or using default credentials. Similarly, \textit{vulnerabilities in Deployment Scripts}, like command injection or insecure permissions, allow attackers to execute arbitrary commands in deployment environments. \textit{Unpatched Software Components.} Failure to update or patch microservice components used in the deployment process, such as dependencies and third-party libraries, leaves the environment vulnerable.

\subsection{Application-Level Vulnerabilities}
Application-related vulnerabilities exist within the individual microservices, irrespective of the underlying orchestration platform. Unlike infrastructure-level vulnerabilities, which focus on weaknesses in the deployment environment or orchestration platform, application-level vulnerabilities pertain specifically to the functionality, implementation, and security of the microservices code, configurations, or runtime environments \cite{yarygina2018overcoming}.

\textbf{Microservice Development Vulnerabilities.} Microservice development vulnerabilities refer to coding errors and security flaws present in the codebase or implementation of individual microservices that can lead to security vulnerabilities. These errors arise due to oversight, misunderstanding of security principles, or incorrect implementation of security measures during the development process. \textit{Input Validation and Sanitization Issues.} Failure to properly validate and sanitize user input leads to vulnerabilities such as SQL injection, cross-site scripting (XSS), or command injection. This allows attackers to manipulate the behavior of microservices or execute unauthorized actions. \textit{Business Logic Vulnerabilities.} Flaws in the business logic of microservices leads to security vulnerabilities. For example, insufficient validation of transactional operations or improper handling of edge cases results in unauthorized actions or unexpected behavior. \textit{Authentication and Authorization Weaknesses.} Weaknesses in authentication mechanisms or insufficient authorization controls can result in unauthorized access to microservices or sensitive data \cite{nkomo2019software}. For example, using weak passwords or lacking proper session management can compromise user authentication.

\textbf{Microservice Communication Vulnerabilities.} Vulnerabilities in the interactions and data exchange between microservices within a distributed system are known as microservice communication vulnerabilities \cite{yarygina2018overcoming}. These vulnerabilities can occur due to insecure communication protocols, insufficient authentication and authorization mechanisms, or inadequate encryption of data transmitted between microservices \cite{yu2019survey}. \textit{Insecure Communication Protocols.} Reliance on insecure communication protocols, such as HTTP without Transport Layer Security (TLS) or unencrypted messaging formats like plain text, can expose sensitive data transmitted between microservices to interception and tampering by attackers \cite{torkura2017integrating}. \text{Service Discovery Vulnerabilities.} Inadequate authentication and validation mechanisms in service discovery protocols may enable attackers to spoof or impersonate legitimate services. This could lead to unauthorized access, data exfiltration, or disruption of service communication \cite{yu2019survey}. \textit{Dependency Chain Risks.} Services may have dependencies on other services or external components (e.g., databases, APIs). Vulnerabilities in upstream dependencies can propagate downstream, affecting multiple services and increasing the attack surface of the system \cite{torkura2017integrating}.

\begin{figure}[t!]
 \centering
\includegraphics[width=0.8\columnwidth,height=0.8\columnwidth]{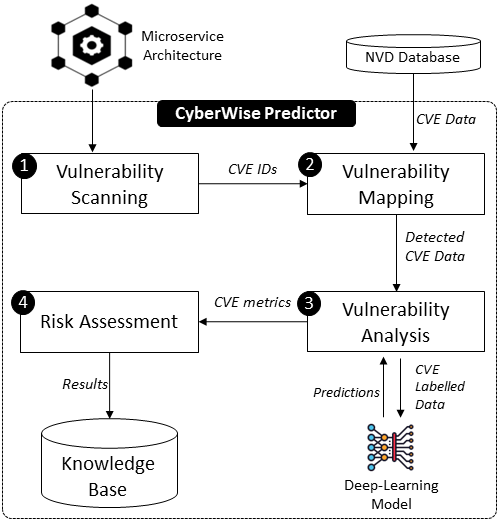}
\caption{CyberWise Predictor architecture.}
 \label{proposedframe}
\end{figure}

\section{CyberWise Predictor Architecture} \label{Section 4}

Figure \ref{proposedframe} illustrates the architecture of \textit{CyberWise Predictor}. \textit{CyberWise Predictor} detects vulnerabilities in microservice architectures using specialized vulnerability scanning tools. Subsequently, it maps the detected vulnerabilities to the National Vulnerability Database (NVD) to retrieve relevant data for each vulnerability. Following this, the system analyzes each vulnerability to calculate cybersecurity risk. For any missing data, a deep learning-based model is employed to predict and assess cybersecurity risks. The results are then stored in a knowledge base for record-keeping. The subsequent sections provide an in-depth exploration of the functionality of each component within \textit{CyberWise Predictor}, accompanied by detailed insights into our experimental setup.

We chose the Kubernetes container orchestration platform for running a benchmark application for microservices, a decision influenced by its widespread adoption in existing literature and related research \cite{kubernetes2019kubernetes}. Our selection of the open-source microservice benchmark application, Sock Shop \cite{microservices-demo}, was driven by its compatibility with Kubernetes and its prevalent use in the current body of literature \cite{van2023continuous, nobre2023anomaly}. This application served as the foundational basis for our service-by-service testing.

\subsection{Phase 1: Vulnerability Scanning}

Vulnerability scanning is the process of identifying security vulnerabilities in microservice architectures. We utilize Trivy\footnote{https://trivy.dev}, an open-source vulnerability scanner, for detecting vulnerabilities in Sock Shop. We first integrate Trivy into \textit{CyberWise Predictor} to conduct scans on microservice images of Sock Shop within the Kubernetes environment. Trivy identifies vulnerabilities, provides information on their corresponding CVE numbers, and stores this data in a CSV file. A total of 2395 vulnerabilities were detected in the Sock Shop. The specifics of the vulnerability scanning results can be found in our replication package \cite{selfadaptive}.

\begin{table}[t!]
\caption{CVSS v2 metric's labels.}
\renewcommand{\arraystretch}{1.1}
\label{cvss labels}
\centering
\begin{tabular}{ll}
\hline
\textbf{Metric} & \textbf{Possible Values} \\ \hline
Attack Vector & Network, Local, Adjacent Network \\ \hline
Attack Complexity & Low, Medium, High \\ \hline
Confidentiality Impact & Complete, Partial, None \\ \hline
Authentication & Multiple, Single, None \\ \hline
Availability Impact & Complete, Partial, None \\ \hline
Integrity Impact & Complete, Partial, None \\ \hline
\end{tabular}
\label{table:cvss_metrics}
\end{table}

\subsection{Phase 2: Vulnerability Mapping}

The \textit{CyberWise Predictor} maps identified vulnerabilities to the ones stored in the National Vulnerability Database (NVD) to extract their metrics, such as probability and impact scores. These metrics are crucial for calculating the cybersecurity risk associated with each vulnerability. We retrieve the CVE vulnerability descriptions, as outlined in Table \ref{cvss labels}. To streamline this mapping process, we adopt a proactive approach by pre-downloading a substantial portion of the NVD, enabling local fetches. This approach minimizes the need for real-time requests to the NVD server, and external connections are made only when updates or new data are required. To extract data, we use the Python library called NVDlib\footnote{https://nvdlib.com/en/latest}, which connects with NVD's API 2.0. From the NVD data, we extract useful information, including vulnerability descriptions and metrics from the CVSS v2. We chose CVSS v2 because it is well-represented in the NVD-CVE dataset, and larger datasets typically lead to improved performance in deep-learning models. It is observed that not all CVEs in the NVD are equipped with CVSS labels, irrespective of the version \cite{inproceedings}. In Table \ref{table1}, the count of detected CVEs with and without CVSS v2 scores is presented for the Sock Shop. The data reveals that 23.96\% of detected vulnerabilities lack CVSS metric scores in the NVD.

\begin{table}[t]
\caption{CVSS v2 Metrics in Sock Shop Vulnerabilities.}
\renewcommand{\arraystretch}{1.1}
\label{table1}
\centering
\begin{tabular}{p{2cm}p{1.5cm}p{1.5cm}p{2cm}}
\hline
\textbf{Sock Shop} & \textbf{Available Metrics} & \textbf{Unavailable Metrics} & \textbf{\% Unavailable Metrics} \\ \hline
Detected CVEs & 1821 & 574 & 23.96\% \\ \hline
\end{tabular}
\label{table:cvss_v2}
\end{table}

\subsection{Phase 3: Vulnerability Analysis} 
In this phase, we categorize the identified vulnerabilities within the Sock Shop application. We examine whether the relevant data associated with the detected vulnerabilities, such as CVSS score and exploitability score, is available in NVD. This is essential for the subsequent calculation of the cybersecurity risks associated with these vulnerabilities. Our categorization is based on the completeness of the CVSS version 2 data for each identified vulnerability. This is done by assigning binary tags (`0' or `1'.) to NVD records: `0' signifies the absence of these metrics, while `1' indicates the presence of metrics. This categorization divides the data into two groups: one with complete CVSS v2 metrics and another without this information. Any empty fields in the dataset are filled with "NF" (Not Found) to handle missing information. The purpose of this categorization is to format the data as input for the subsequent deep learning prediction model in the next stage.\\

\noindent \textbf{Addressing Missing CVSS data with Deep Learning.} To address the absence of CVSS scores for identified vulnerabilities, we employ Natural Language Processing (NLP) models based on deep learning frameworks to predict missing CVSS metrics using vulnerability descriptions. Initially, we aggregate data from the NVD-CVSS v2 database to fine-tune our NLP model, enabling it to predict all six CVSS v2 metrics (e.g., access vector and access complexity) based on vulnerability descriptions.  Subsequently, we explore the NLP techniques employed and the challenges faced in developing this model. This includes addressing imbalanced training data, selecting high-performance models, and overcoming hardware limitations in training deep-learning models.\\

\begin{figure}[t]
  \centering
  \includegraphics[width=0.95\columnwidth]{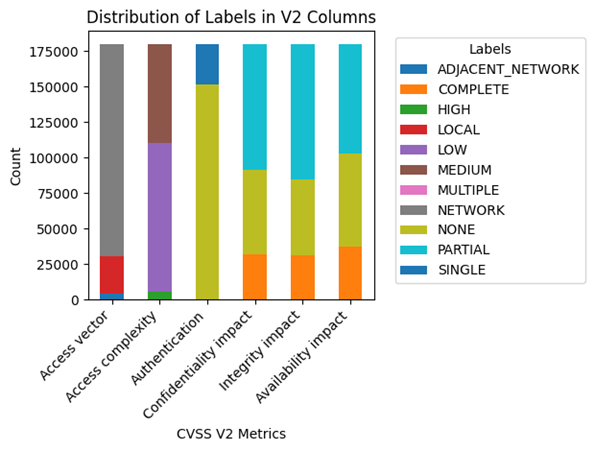}
  \caption{CVSS v2 labels imbalances.}
  \label{lblimb}
\end{figure}

\noindent \textbf{Addressing Imbalanced Data through Weighted Label Approach.}
After data analysis, our methodology performs multi-label classification, where each of the six CVSS v2 metrics has three possible values. Consequently, we consider each metric as an independent classification problem. The vulnerability description serves as the model input, while the metric labels serve as the outputs. This approach results in one of three potential labels for each metric. Upon visualizing the distribution of these metric labels, we observed an imbalance (as shown in Figure \ref{lblimb}). Such imbalance may cause the model to prioritize predicting the majority classes, potentially neglecting the minority ones. To mitigate this issue, oversampling of minority classes is a potential solution. However, oversampling must be performed cautiously to prevent introducing bias into the model. To address this imbalance, \textit{we implement a weighted label approach} \cite{wu2018weighted, aurelio2019learning}. We ensure each label in the dataset contributes appropriately to the model's training process by employing a weighted loss technique, as discussed below.

\begin{enumerate}
    \item We first calculate the frequency of each label by counting the samples for each label, denoted as label $i$. The frequency $F_i$ is determined as follows:
    \[ F_i = \text{count}(\text{samples of label } i) \]

    \item We then compute label weights, where the weight for each label $w_i$ is obtained by dividing the total number of samples by the frequency of the label:
    \[ w_i = \frac{N}{F_i} \]
   where $N$ denotes the total number of samples.
    
    \item We integrate class weights into the loss function by adjusting the standard cross-entropy loss function. The standard cross-entropy loss function is given by:
    \[ \text{loss} = -\sum(\text{true} \times \log(\text{prediction})) \]
    
    The modified loss function incorporating class weights is computed as:
    \[ \text{loss} = -\sum(\text{true} \times w_i \times \log(\text{prediction}_i)) \]
\end{enumerate}

\noindent \textbf{Example. } If there are 100 samples in total, with 90 of them labeled as "A" and 10 as "B".

\[
\text{A-weight} = \frac{100}{90} = 1.11
\]
\[
\text{B-weight} = \frac{100}{10} = 10
\]

The values indicate that class B errors have a tenfold greater impact on the overall loss compared to class A errors. In this way, we aim to mitigate the under-representation of class B in the dataset, ensuring that the model adequately learns from this class during training. We divided the dataset into 80\% for training, with 10\% set aside for validation and another 10\% for testing purposes. Table \ref{traintest} illustrates the distribution of instances across the training, validation, and testing sets. Preserving the integrity of the evaluation and testing datasets is crucial to ensure the reliability of the results. The validation dataset is utilized to assess model performance after each training epoch, aiding in the detection and prevention of over-fitting, where the model simply memorizes the training data.\\

\begin{table}[t]
\caption{Data split into training, testing, and validation sets.}
\renewcommand{\arraystretch}{1.1}
\centering
\label{traintest}
\begin{tabular}{p{2.5cm}p{3.5cm}}
\hline
\textbf{Section} & \textbf{Number of records} \\ \hline
Training & 141392 \\ \hline
Validation & 17674 \\ \hline
Testing & 17674 \\ \hline
\end{tabular}
\label{table:dataset}
\end{table}

\begin{figure}[b]
  \centering
  \includegraphics[width=0.9\columnwidth]{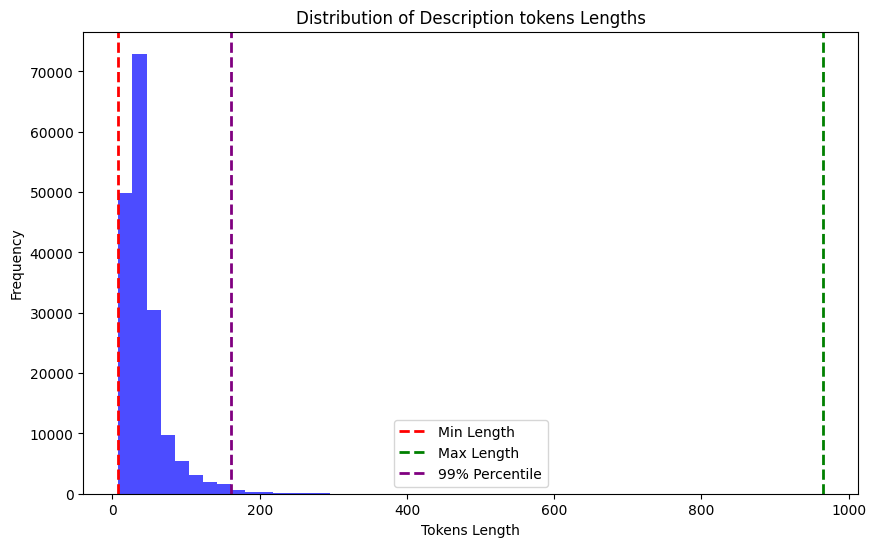}
  \caption{Distribution of CVSS availability.}
  \label{distoOflength}
\end{figure}

\noindent \textbf{Model Selection and Resource Considerations in Deep Learning Transformer Architectures.}
During the model selection process for our experiments, we analyzed various transformer models from Hugging Face \cite{wolf2019huggingface} and Google \cite{tunstall2022natural}. Both BERT \cite{devlin2018bert} and XLNet \cite{yang2019xlnet} initially appeared promising, but their resource demands and memory consumption posed significant challenges. BERT, known for its size, and XLNet, designed for longer inputs, presented difficulties during training and tuning. Through our analysis, RoBERTa \cite{liu2019roberta} emerged as a model that aligned well with our objectives. RoBERTa, a compromise between BERT and XLNet, exhibits good convergence speed with lower memory consumption compared to XLNet, and achieves high accuracy (Section \ref{Section 5.1}).  Derived from BERT, RoBERTa maintained efficacy while being more manageable in size and training complexity. Our analysis highlighted the importance of careful parameter selection, given the resource-intensive nature of these models. We found that training on a GPU with 12 gigabytes of memory required meticulous attention, as exceeding this capacity risked data spillage into slower system memory, resulting in substantial delays in data transfers between the system and GPU onboard memory.

We analyzed the impact of input length on model performance, which proved to be crucial. Choosing an excessively large input size risked memory overspill, while a smaller size led to input truncation. To strike a balance between length and contextual information, we conducted an analysis where we removed stop words to shorten the input without sacrificing context. Our analysis of tokenized descriptions revealed that 90\% of descriptions remained within a 130-token length after processing, as illustrated in Figure \ref{distoOflength}. Inputs shorter than 130 tokens resulted in decreased model accuracy due to the reduced amount of data provided. We extended this length by two special tokens, resulting in an input length of 132 tokens. It's worth noting that, without significant resource constraints, we determined the optimal input maximum length to cover 99\% of tokenized text, which is 256 tokens.

 \begin{figure}[t]
  \centering
  \includegraphics[width=0.95\columnwidth]{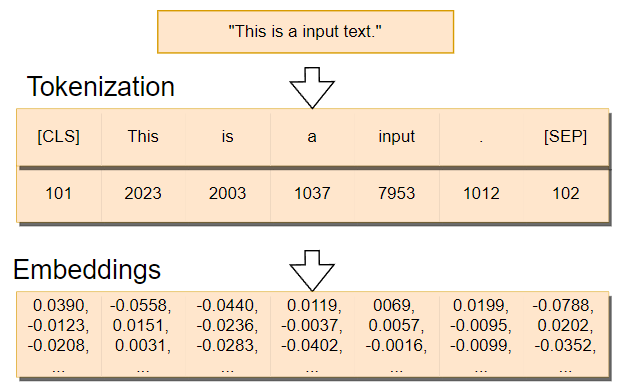}
  \caption{Tokenization of text.}
  \label{token}
\end{figure}

Additionally, we conducted analysis to determine the appropriate batch size for training, considering two critical factors: training time and model tuning. Smaller batch sizes were found to prolong training duration due to more frequent updates. Since each metric was treated as a distinct classification task, the number of epochs varied across training sessions. To pinpoint the optimal stopping point for training each metric, we employed an early stopping technique. Instead of adhering to a fixed number of epochs, this adaptive strategy dynamically adjusted the training duration, optimizing performance for each specific classification challenge. This flexible approach accommodated the unique characteristics of each classification task, ultimately enhancing the accuracy of predictions. Figure \ref{token} presents the tokenization of text.\\

\begin{figure}[t]
  \centering
  \includegraphics[width=0.9\columnwidth]{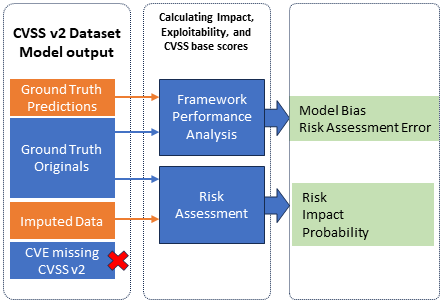}
  \caption{Data used in model evaluation.}
  \label{dataflowdig}
\end{figure}

\noindent \textbf{Integrating Ground Truth in Risk Assessment Methodology.}
After data mapping, which involved integrating a new column into our dataset. This additional column served as a label, where we assigned `1' to data entries containing complete CVSS v2 metrics (considered as ground truth), and `0' to entries lacking such metrics. This labeling approach was crucial, as it allowed us to differentiate between complete and incomplete data for all CVEs, which were subject to prediction of their CVSS v2 metrics by our model. The data labeled with `1' served as the ground truth, serving as a reference for error measurement within our model and framework. This ground truth, alongside the imputation of missing data, played a pivotal role in our risk assessment process. We illustrated the significance of both the established ground truth and the imputation of missing data in Figure \ref{dataflowdig}, highlighting their importance in our risk assessment methodology.

\subsection{Phase 4: Risk Assessment}

In our risk assessment process, we merge the CVSS ground truth with our model predictions. When dealing with records containing known CVSS v2 metrics, we directly utilize the actual values. However, for instances where the CVSS v2 metrics are unavailable, we employ our model's predictions to fill in the missing data. This dual approach ensures that our risk evaluation process incorporates both the established ground truth and the insights generated by our model, thereby facilitating a comprehensive and precise assessment of the risk associated with each CVE record. To quantify the risk associated with each CVE record, we calculate the impact score, exploitability score, and base score using the following equations,  as described in further detail in \cite{Equations}.

\begin{equation*}
    \begin{aligned}
        \text{Base Score} = \left( (0.6 \times \text{Impact Score} + 0.4 \times \text{Exploitability Score} \right. \\
        \left. - 1.5) \times f(\text{Impact}) \right)
    \end{aligned}
\end{equation*}

\begin{align*}
    \begin{split}
    \text{Impact Score} = 10.41 \times (1 - (1 - \text{Confidentiality Impact})\\ 
    \times (1 - \text{Integrity Impact}) \times (1 - \text{Availability Impact}))
    \end{split}
\end{align*}

\begin{align*}
    \begin{split}
    \text{Exploitability Score} = 20 \times \text{AccessVector} \times \\ \text{AccessComplexity} \times \text{Authentication}
    \end{split}
\end{align*}

\begin{align*}
f(Impact) = \begin{cases}
0 & \text{if Impact = 0} \\
1.176 & \text{otherwise}
\end{cases}
\end{align*}

\section{Experimental Results} \label{Section 5}
This section presents experimental results on fine-tuning transformer models, namely BERT \cite{devlin2018bert}, RoBERTa \cite{liu2019roberta}, and XLNet \cite{yang2019xlnet}, using the NVD-CVE CVSS version 2 dataset. Among these models, RoBERTa emerges as the best performer, demonstrating precision in predicting CVSS metrics on the test set. We further assess the impact of CVSS version 2 data imputation on risk assessment, utilizing the Sock Shop microservice as a case study within our \textit{CyberWise Predictor}. The details of the transformer model fine-tuning implementation are documented in our replication package \cite{finetuner}. Our implementation of RoBERTa provides flexibility to enable the interchange of models and adjustment of parameters, thereby facilitating further investigation and experimentation. In addition, details of the \textit{CyberWise Predictor} can be found in \cite{selfadaptive}.

\begin{table}[t]
\caption{Parameters of each metric classification.}
\renewcommand{\arraystretch}{1.2}
\label{parameterspecs}
\begin{tabular}{lp{0.9cm}p{1.9cm}p{0.8cm}p{0.6cm}}
\hline
\textbf{Metric} & \textbf{Epochs} & \textbf{Input Length} & \textbf{Batch} \\ \hline
Access Vector & 10 & 256 & 128 \\ \hline
Access Complexity & 10 & 256 & 128 \\ \hline
Authentication & 20 & 256 & 128 \\ \hline
Confidentiality Impact & 20 & 256 & 64 \\ \hline
Integrity Impact & 24 & 256 & 64 \\ \hline
Availability Impact & 24 & 256 & 64 \\ \hline
\end{tabular}
\end{table}

\begin{table*}[t]
\centering
\caption{Characteristics comparison of BERT, RoBERTa, and XLNet transformer models.}
\renewcommand{\arraystretch}{1.1}
\label{Bertvsberta}
\begin{tabular}{p{5cm}p{3cm}p{3cm}p{2cm}}
\hline
\textbf{Feature} & \textbf{BERT} & \textbf{RoBERTa} & \textbf{XLNet} \\ \hline
Memory requirements & Up to 1.3 GB & Up to 1.5 GB & Up to 3.5 GB \\ \hline
Speed of tuning on custom dataset & Slow & Fast & Fast \\ \hline
Max input size & 512 tokens & 512 tokens & 2048 tokens \\ \hline
Classification performance & Good & Excellent & N/A \\ \hline
Number of parameters & 110M & 125M & 340M \\ \hline
\end{tabular}%
\end{table*}

\begin{table}[t]
\caption{Accuracy comparison between BERT and RoBERTa.}
\renewcommand{\arraystretch}{1.1}
\centering
\begin{tabular}{lcc}
\hline
\textbf{Metric} & \textbf{BERT} & \textbf{RoBERTa}  \\ \hline
Access Vector & 0.901 & 0.944  \\ \hline
Access Complexity & 0.882 & 0.928 \\ \hline
Authentication &0.89 & 0.936  \\ \hline
Confidentiality Impact &0.86 & 0.92  \\ \hline
Integrity Impact &0.86 & 0.912  \\ \hline
Availability Impact &0.84 & 0.896  \\ \hline
\end{tabular}%
\label{bvb2}
\end{table}

\subsection{Results of Deep Learning Fine-Tuning} \label{Section 5.1}

The primary objective of this work is to fine-tune a deep learning transformer model using the NVD-CVE CVSS version 2 dataset, which comprises vulnerability descriptions. Our aim is to predict the CVSS metrics (Access Vector, Access Complexity, Authentication, Confidentiality Impact, Integrity Impact, and Availability Impact) for each vulnerability, serving as the model's outputs. Table \ref{parameterspecs} presents the parameters utilized for the classification of each CVSS metric, including the number of epochs, learning rate, input length, and batch size.

Moreover, we conducted experiments using three transformer models: BERT, RoBERTa, and XLNet. Initially, we selected BERT, which exhibited an average predictive accuracy of 88\% across the six classification tasks. However, the training time BERT extended to approximately 20 minutes per epoch, which is too high. Consequently, we tried XLNet, which showed faster convergence than BERT but faced challenges with slow convergence due to the requirement to start with a small learning rate. Despite using more than double the memory of BERT, attempts to alleviate memory usage by reducing the batch size resulted in significantly slower training speed. As a result, we decided to abandon XLNet without achieving substantial results.

Finally, RoBERTa emerged as a balanced compromise between BERT and XLNet. RoBERTa demonstrated satisfactory convergence speed, with memory consumption remaining lower than that of XLNet. The results from RoBERTa indicate high accuracy. Table \ref{Bertvsberta} summarizes essential characteristics of BERT, RoBERTa, and XLNet, including memory requirements, tuning speed on custom datasets, maximum input size, classification performance, and number of parameters. Additionally, Table \ref{bvb2} provides the accuracy comparison between BERT and RoBERTa, focusing on key metrics. The results highlight the superior performance of RoBERTa across multiple dimensions in our experimental setup. We evaluated RoBERTa outcomes using these metrics:

\begin{itemize}[leftmargin=*]
    \item \textbf{Accuracy.} Accuracy denotes the correctness of a model's predictions, reflecting the proportion of correctly classified instances.
    \item \textbf{Recall.} Recall is defined as the model's ability to identify and capture all relevant instances in a dataset, emphasizing the completeness of its output.
    \item \textbf{Precision.} Precision is a performance metric in machine learning that represents the accuracy of positive predictions by measuring the ratio of true positive predictions to the sum of true positives and false positives.
    \item \textbf{F1-Weighted Score.}  The F1-weighted score is a metric that combines precision and recall using the harmonic mean. It provides a balanced measure of a model's performance, particularly suitable for imbalanced datasets, by considering both false positives and false negatives.

    \item \textbf{F1-Micro Score.} Micro F1 score calculates the overall F1 score by considering all classes together, regardless of class imbalances. It is computed by aggregating the true positives, false positives, and false negatives across all classes and then calculating the F1 score, providing a holistic measure of a model's performance across the entire dataset.
    
\end{itemize}

\begin{table*}[t!]
\caption{Performance metrics of fine-tuned RoBERTa model on the test dataset for CVSS metrics.}
\label{testsetresults}
\renewcommand{\arraystretch}{1.1}
\begin{tabular}{lccccc}
\hline
\textbf{CVSS Metric} & \textbf{Accuracy} & \textbf{F1 Score (Weighted)} & \textbf{F1 Score (Micro)} & \textbf{Recall} & \textbf{Precision}\\ \hline
Access Vector & 0.944 & 0.944 & 0.944 & 0.944 & 0.945 \\ \hline
Access Complexity & 0.928 & 0.926 & 0.928 & 0.928 & 0.934 \\ \hline
Authentication & 0.936 & 0.936 & 0.936 & 0.936 & 0.938 \\ \hline
Confidentiality Impact & 0.92 & 0.913 & 0.929 & 0.92 & 0.921 \\ \hline
Integrity Impact & 0.912 & 0.912 & 0.912 & 0.912 & 0.912 \\ \hline
Availability Impact & 0.896 & 0.895 & 0.896 & 0.896 & 0.896 \\ \hline
\end{tabular}%
\end{table*}

The results of fine-tuning the RoBERTa transformer model exhibit a high accuracy rate of \textbf{92\%} on the test set. This is because the RoBERTa model was first trained on a large amount of English language data. This initial training helped RoBERTa understand the subtle meanings and details in written text more deeply. Our process of pre-training the model and then fine-tuning it has significantly improved its overall performance. In Table \ref{testsetresults}, we present test set results obtained from the fine-tuned RoBERTa transformer model. The table shows various metrics, including Accuracy, F1 Score, Recall, and Precision, across different CVSS metrics. The fine-tuned RoBERTa model archives the best performance on the test set, with accuracy values ranging from 0.896 to 0.944, averaging at 0.922. Notably, it exhibited consistent F1 scores, recall, and precision across other key attributes. These results underscore the model's robust classification capabilities.

Furthermore, we utilized the F1 score as a metric to evaluate the fairness of our model's predictions. Despite potential imbalances in the labels, we found that incorporating label weights into the loss function, as indicated by the F1 score, effectively helped mitigate bias in the model's predictions, aligning with our desired objective.

\begin{table*}[t]
\centering
\caption{Error percentages in impact, exploitability, and base scores for each microservice in the Sock Shop.}
\renewcommand{\arraystretch}{1.1}
\label{errorresults}
\begin{tabular}{lccc}
\hline
\textbf{Component} & \textbf{Impact Score Error} & \textbf{Exploitability Score} \textbf{Error} & \textbf{Base Score Error} \\ \hline
carts & 0.97 & 0.144 & 0.402 \\ \hline
carts-db & 0.868 & 0.556 & 0.601 \\ \hline
catalogue & 1.186 & 0.147 & 0.462 \\ \hline
catalogue-db & 0.86 & 0.509 & 0.549 \\ \hline
front-end & 1.022 & 0.141 & 0.562 \\ \hline
orders & 0.97 & 0.144 & 0.402 \\ \hline
orders-db & 0.868 & 0.556 & 0.601 \\ \hline
payment & 1.186 & 0.147 & 0.462 \\ \hline
queue-master & 0.991 & 0.149 & 0.454 \\ \hline
rabbitmq & 0.906 & 0.498 & 0.558 \\ \hline
shipping & 0.978 & 0.147 & 0.41 \\ \hline
user & 1.284 & 0.129 & 0.621 \\ \hline
user-db & 0.883 & 0.515 & 0.567 \\ \hline
\end{tabular}%
\end{table*}

\begin{table*}[t]
\centering
\renewcommand{\arraystretch}{1.1}
\caption{Bias of prediction-driven scores. Lower absolute values indicate better performance.}
\label{biastable}
\begin{tabular}{lccc}
\hline
\textbf{Component} & \textbf{Impact Score Bias} &\textbf{ Exploitability Score Bias} & \textbf{Base Score Bias} \\ \hline
carts & -0.003 & 0.005 & 0.015 \\ \hline
carts-db & 0.004 & -0.017 & 0.005 \\ \hline
catalogue & 0.024 & 0.012 & 0.014 \\ \hline
catalogue-db & 0.001 & -0.011 & 0.003 \\ \hline
front-end & 0.005 & 0.009 & 0.019 \\ \hline
orders & -0.003 & 0.005 & 0.015 \\ \hline
orders-db & 0.004 & -0.017 & 0.005 \\ \hline
payment & 0.024 & 0.012 & 0.014 \\ \hline
queue-master & 0.0001 & 0.005 & 0.015 \\ \hline
rabbitmq & 0.005 & -0.01 & 0.005 \\ \hline
shipping & -0.002 & 0.005 & 0.016 \\ \hline
user & 0.037 & 0.009 & 0.018 \\ \hline
user-db & 0.004 & -0.011 & 0.004 \\ \hline
\end{tabular}%
\end{table*}

\subsection{Results of the Risk Assessment}
Within our self-adaptive framework, we subjected the microservice to a comprehensive evaluation, including vulnerability detection, mapping to the NVD database, and imputation using a pre-trained deep learning model. Subsequently, we calculated impact, exploitability, and base scores, which are vital indicators of vulnerability severity. To compute the effectiveness of CVSS version 2 data imputation, we used two measures. Firstly, we utilized the \textit{percentage error}, which evaluates the accuracy of a score by comparing our predictions with actual scores. Secondly, we used the concept of \textit{bias}, representing the deviation between actual and predicted values. These metrics were defined by Equation \ref{percent_error} and Equation \ref{bias}, respectively. Analyzing bias allowed us to discern whether our process tended to overestimate or underestimate scores and, consequently, the associated risk. We observed that the bias value offered valuable insights into the extent of deviation between our predictions and the actual CVSS version 2 scores, which range from 0 to 10. A consistently positive or negative bias in our results indicated whether our process tended to systematically overemphasize or underestimate the severity of vulnerabilities.

\begin{equation} \label{percent_error}
\text{Percentage error} = \left| \frac{\text{true value} - \text{predict value}}{\text{true value}} \right| \times 100
\end{equation}

\begin{equation} \label{bias}
\text{Bias} = \text{true value} - \text{predict value}
\end{equation}

Table \ref{errorresults} outlines the error percentages for impact scores, exploitability scores, and base scores of each service in the Sock Shop dataset. Our results underscore the accuracy of our prediction-driven scoring method, with minimal errors observed when comparing predicted and actual values for each service. This highlights the reliability of our deep learning model in estimating missing Common Vulnerability Scoring System (CVSS) version 2 metrics. Furthermore, the consistency across various services indicates the robustness of our approach in handling diverse components within the dataset. These findings affirm the efficacy of employing predictive modeling techniques to fill in the gaps in vulnerability scoring, thereby enhancing the overall security assessment process and offering valuable insights into the reliability of CVSS version 2 metric estimations for a broad spectrum of services.

Additionally, Table \ref{biastable} presents the Bias of prediction-driven scores. Our bias analysis reveals minimal biases in our research results, predominantly in underestimating risk scores. The consistently small errors across the table underscore the reliability and precision of our approach, laying a robust foundation for its practical application in real-world vulnerability risk assessments. The bias analysis reveals minimal biases, primarily underestimating risk scores, indicating the reliability of our approach in predicting vulnerability scores.

\section{Discussion} \label{Section 6}

\textbf{Comparison of our study with existing studies.} This work highlights the effectiveness of RoBERTa, a DL-based NLP model, in the fine-tuning process for imputing CVSS version 2 metrics. Unlike previous studies \cite{sparck1972statistical, mikolov2013distributed} that often rely on traditional machine learning approaches, we focused on advanced DL models aiming to enhance the precision of cybersecurity risk prediction in microservices. Our experimental results show that with the transformer-based language model, we achieved promising results, with improved accuracy scores averaging at \textbf{92\%}, surpassing the performance reported in other studies \cite{RN3, yang2019xlnet}. A key aspect of our experiments that contributed to minimizing errors is the collaborative use of ground truth alongside predicted data. This approach helped us achieve superior accuracy in CVSS version 2 imputation for microservices. Our work shows RoBERTa as the optimal model, as it outperformed alternatives like BERT and XLNet. Our findings not only underscore the practical applicability of RoBERTa but also 
diverges from and extends upon existing literature by employing advanced NLP and deep learning models for vulnerability assessment, a domain that, until now, had seen limited application of these sophisticated AI techniques \cite{dragoni2017microservices}.

\textbf{Implications of our study for researchers.} 
We specifically focus on utilizing professional vulnerability descriptions sourced from the National Vulnerability Database for both training and testing purposes. Further investigations could explore the potential of our framework in predicting user-reported descriptions and descriptions sourced from diverse databases and platforms. Additionally, besides formal vulnerability repositories like NVD, we suggest exploring informal sources such as technical forums like Stack Overflow for enhancing the security assessment of microservice architectures. 
Moreover, it is worthwhile to delve into the utilization of reinforcement learning models, along with examining the extent to which the NLP-based prediction model depends on the professionalism and consistency of vulnerability sources.

\textbf{Implications of our study for practitioners.} 
Our findings emphasize the significance of CVSS data imputation for improving vulnerability prediction and assessment. Similar to existing literature, we acknowledge the challenge presented by incomplete CVSS labels within the NVD dataset \cite{9513723}. Our research has direct implications for cybersecurity professionals, software developers, and organizations employing microservice architecture. Through accurate imputation of missing CVSS data, our framework elevates risk assessment capabilities, aiding in the prediction of vulnerabilities and facilitating effective assessment strategies. Our study acts as a bridge between academia and industry, providing practical tools to enhance cybersecurity practices. The findings contribute to the broader research landscape by highlighting the superiority of RoBERTa for CVSS data imputation. This insight guides practitioners in selecting appropriate deep-learning models for similar tasks, improving both accuracy and efficiency. Furthermore, this study enriches the academic discourse, offering a foundational framework that underscores the utility of deep learning and NLP in cybersecurity, thus paving the way for future investigations into their broader application across various cybersecurity challenges. 

\textbf{Limitations of our study.} While our RoBERTa model demonstrated high capability and accuracy, we encountered challenges with some vulnerability descriptions having an insufficient word count (less than 8 words). This limitation prevented the model from deriving meaningful predictions due to the lack of descriptive text. Although these cases were relatively low in frequency and did not significantly impact the overall experiment, they did limit the model's potential for achieving higher performance. Additionally, the fine-tuning process proved to be time-consuming, primarily due to the complexity of the language models. Each run demanded hours to complete, despite leveraging accelerated computing and mixed float-point calculations \cite{nvidia2023}. This extended duration highlights the intricate nature of the fine-tuning task and the computational demands imposed by advanced language models.

\section{Conclusion and Future Work} \label{Section 7}

We propose the \textit{CyberWise Predictor}, a framework designed for predicting and assessing cybersecurity risks inherent in microservice architectures. To tackle the challenge of unavailable vulnerability data in NVD, we leverage RoBERTa, a transformer-based language model fine-tuned for predicting missing CVSS metrics. Our results underscore the effectiveness of \textit{CyberWise Predictor}, achieving an average accuracy of 92\% in predicting vulnerabilities within a microservice benchmark application (Sock Shop). Furthermore, our risk assessment framework demonstrates reliability by computing risk scores with minimal error and bias. Additionally, we introduce a taxonomy for vulnerabilities associated with microservices.

In the future, we aim to evaluate the effectiveness of \textit{CyberWise Predictor} not only with formal vulnerability repositories like NVD but also by exploring informal sources such as technical forums such as Stack Overflow to enhance the security assessment of microservice architectures. Furthermore, it would be valuable to explore the utilization of reinforcement learning models to predict missing risk assessment metrics. In addition, we plan to expand our research to assess how \textit{CyberWise Predictor} predicts vulnerability scores when defense mechanisms are present and how these defense mechanisms influence changes in vulnerability scores.

\bibliographystyle{unsrt}
\bibliography{ref}

\begin{thebibliography}{10}

\bibitem{pimentel2021self}
Eliaquim Pimentel, Wellington Pereira, Paulo Henrique~M Maia, Mariela~I Cort{\'e}s, et~al.
\newblock Self-adaptive microservice-based systems-landscape and research opportunities.
\newblock In {\em International Symposium on Software Engineering for Adaptive and Self-Managing Systems (SEAMS)}, pages 167--178. IEEE, 2021.

\bibitem{rossi2022dynamic}
Fabiana Rossi, Valeria Cardellini, Francesco~Lo Presti, and Matteo Nardelli.
\newblock Dynamic multi-metric thresholds for scaling applications using reinforcement learning.
\newblock {\em IEEE Transactions on Cloud Computing}, 2022.

\bibitem{ahmad2023review}
Hussain Ahmad, Isuru Dharmadasa, Faheem Ullah, and Muhammad~Ali Babar.
\newblock A review on c3i systems’ security: Vulnerabilities, attacks, and countermeasures.
\newblock {\em ACM Computing Surveys}, 55(9):1--38, 2023.

\bibitem{ahmad2024smart}
Hussain Ahmad, Christoph Treude, Markus Wagner, and Claudia Szabo.
\newblock Smart hpa: A resource-efficient horizontal pod auto-scaler for microservice architectures.
\newblock {\em arXiv preprint arXiv:2403.07909}, 2024.

\bibitem{blinowski2022monolithic}
Grzegorz Blinowski, Anna Ojdowska, and Adam Przyby{\l}ek.
\newblock Monolithic vs. microservice architecture: A performance and scalability evaluation.
\newblock {\em IEEE Access}, 10:20357--20374, 2022.

\bibitem{bushong2021microservice}
Vincent Bushong, Amr~S Abdelfattah, Abdullah~A Maruf, Dipta Das, Austin Lehman, Eric Jaroszewski, Michael Coffey, Tomas Cerny, Karel Frajtak, Pavel Tisnovsky, et~al.
\newblock On microservice analysis and architecture evolution: A systematic mapping study.
\newblock {\em Applied Sciences}, 11(17):7856, 2021.

\bibitem{mateus2021security}
Nuno Mateus-Coelho, Manuela Cruz-Cunha, and Luis~Gonzaga Ferreira.
\newblock Security in microservices architectures.
\newblock {\em Procedia Computer Science}, 181:1225--1236, 2021.

\bibitem{dragoni2017microservices}
Nicola Dragoni, Saverio Giallorenzo, Alberto~Lluch Lafuente, Manuel Mazzara, Fabrizio Montesi, Ruslan Mustafin, and Larisa Safina.
\newblock Microservices: yesterday, today, and tomorrow.
\newblock {\em Present and ulterior software engineering}, pages 195--216, 2017.

\bibitem{goel2018overview}
Diksha Goel and Ankit~Kumar Jain.
\newblock Overview of smartphone security: Attack and defense techniques.
\newblock In {\em Computer and Cyber Security}, pages 249--279. Auerbach Publications, 2018.

\bibitem{chandramouli2019microservices}
Ramaswamy Chandramouli.
\newblock Microservices-based application systems.
\newblock {\em NIST Special Publication}, 800(204):800--204, 2019.

\bibitem{montesi2018decorator}
Fabrizio Montesi and Janine Weber.
\newblock From the decorator pattern to circuit breakers in microservices.
\newblock In {\em Proceedings of the 33rd Annual ACM Symposium on Applied Computing}, pages 1733--1735, 2018.

\bibitem{he2017authentication}
Xiuyu He and Xudong Yang.
\newblock Authentication and authorization of end user in microservice architecture.
\newblock In {\em Journal of Physics: Conference Series}, volume 910, page 012060. IOP Publishing, 2017.

\bibitem{RN11}
Kennedy~A Torkura, Muhammad~IH Sukmana, Feng Cheng, and Christoph Meinel.
\newblock Cavas: Neutralizing application and container security vulnerabilities in the cloud native era.
\newblock In {\em Security and Privacy in Communication Networks: 14th International Conference, SecureComm 2018, Singapore, Singapore, August 8-10, 2018, Proceedings, Part I}, pages 471--490. Springer, 2018.

\bibitem{CVSS_Score}
Vulnerability~Metrics National Vulnerability~Database.
\newblock \url{https://nvd.nist.gov/vuln-metrics/cvss#}, Last Access: Dec 2023.

\bibitem{NVD_Database}
General~Information National Vulnerability~Database.
\newblock \url{https://nvd.nist.gov/general}, Last Access: Dec 2023.

\bibitem{RN3}
Xuanyu Duan, Mengmeng Ge, Triet Huynh~Minh Le, Faheem Ullah, Shang Gao, Xuequan Lu, and M~Ali Babar.
\newblock Automated security assessment for the internet of things.
\newblock In {\em 2021 IEEE 26th Pacific Rim International Symposium on Dependable Computing (PRDC)}, pages 47--56. IEEE, 2021.

\bibitem{abubakar2022sentiment}
Haisal~Dauda Abubakar, Mahmood Umar, and Muhammad~Abdullahi Bakale.
\newblock Sentiment classification: Review of text vectorization methods: Bag of words, tf-idf, word2vec and doc2vec.
\newblock {\em SLU Journal of Science and Technology}, 4(1 \& 2):27--33, 2022.

\bibitem{vegesna2023utilising}
Vinod~Varma Vegesna.
\newblock Utilising vapt technologies (vulnerability assessment \& penetration testing) as a method for actively preventing cyberattacks.
\newblock {\em International Journal of Management, Technology and Engineering}, 12, 2023.

\bibitem{wolf2020transformers}
Thomas Wolf, Lysandre Debut, Victor Sanh, Julien Chaumond, Clement Delangue, Anthony Moi, Pierric Cistac, Tim Rault, R{\'e}mi Louf, Morgan Funtowicz, et~al.
\newblock Transformers: State-of-the-art natural language processing.
\newblock In {\em Proceedings of the 2020 conference on empirical methods in natural language processing: system demonstrations}, pages 38--45, 2020.

\bibitem{wolf2019huggingface}
Thomas Wolf, Lysandre Debut, Victor Sanh, Julien Chaumond, Clement Delangue, Anthony Moi, Pierric Cistac, Tim Rault, R{\'e}mi Louf, Morgan Funtowicz, et~al.
\newblock Huggingface's transformers: State-of-the-art natural language processing.
\newblock {\em arXiv preprint arXiv:1910.03771}, 2019.

\bibitem{selfadaptive}
Replication package~of CyberWise~Predictor.
\newblock \url{https://github.com/MajidAbdulsatar/CyberWise-Predictor}.

\bibitem{finetuner}
CyberWise Predictor~DL Model.
\newblock \url{https://github.com/MajidAbdulsatar/CyberWise-Predictor-DL-model-finetuner}.

\bibitem{shah2014automated}
Sugandh Shah and BM~Mehtre.
\newblock An automated approach to vulnerability assessment and penetration testing using net-nirikshak 1.0.
\newblock In {\em 2014 IEEE International Conference on Advanced Communications, Control and Computing Technologies}, pages 707--712. IEEE, 2014.

\bibitem{blinowski2020cve}
Grzegorz~J Blinowski and Pawe{\l} Piotrowski.
\newblock Cve based classification of vulnerable iot systems.
\newblock In {\em Theory and Applications of Dependable Computer Systems: Proceedings of the Fifteenth International Conference on Dependability of Computer Systems DepCoS-RELCOMEX, June 29--July 3, 2020, Brun{\'o}w, Poland 15}, pages 82--93. Springer, 2020.

\bibitem{guo2005automated}
Fanglu Guo, Yang Yu, and Tzi-cker Chiueh.
\newblock Automated and safe vulnerability assessment.
\newblock In {\em 21st Annual Computer Security Applications Conference (ACSAC'05)}, pages 10--pp. IEEE, 2005.

\bibitem{ge2017framework}
Mengmeng Ge, Jin~B Hong, Walter Guttmann, and Dong~Seong Kim.
\newblock A framework for automating security analysis of the internet of things.
\newblock {\em Journal of Network and Computer Applications}, 83:12--27, 2017.

\bibitem{sahner2012performance}
Robin~A Sahner, Kishor Trivedi, and Antonio Puliafito.
\newblock {\em Performance and reliability analysis of computer systems: an example-based approach using the SHARPE software package}.
\newblock Springer Science \& Business Media, 2012.

\bibitem{RN9}
Umesh~Kumar Singh and Chanchala Joshi.
\newblock Quantitative security risk evaluation using cvss metrics by estimation of frequency and maturity of exploit.
\newblock In {\em Proceedings of the World Congress on Engineering and Computer Science}, volume~1, pages 19--21, 2016.

\bibitem{RN10}
Hasan Cam.
\newblock Risk assessment by dynamic representation of vulnerability, exploitation, and impact.
\newblock In {\em Cyber Sensing 2015}, volume 9458, pages 71--79. SPIE, 2015.

\bibitem{yang2020better}
Heedong Yang, Seungsoo Park, Kangbin Yim, and Manhee Lee.
\newblock Better not to use vulnerability’s reference for exploitability prediction.
\newblock {\em Applied Sciences}, 10(7):2555, 2020.

\bibitem{edkrantz2015predicting}
MICHEL Edkrantz.
\newblock Predicting exploit likelihood for cyber vulnerabilities with machine learning.
\newblock 2015.

\bibitem{bullough2017predicting}
Benjamin~L Bullough, Anna~K Yanchenko, Christopher~L Smith, and Joseph~R Zipkin.
\newblock Predicting exploitation of disclosed software vulnerabilities using open-source data.
\newblock In {\em Proceedings of the 3rd ACM on International Workshop on Security and Privacy Analytics}, pages 45--53, 2017.

\bibitem{nowak2021conversion}
Maciej Nowak, Micha{\l} Walkowski, and S{\l}awomir Sujecki.
\newblock Conversion of cvss base score from 2.0 to 3.1.
\newblock In {\em 2021 International Conference on Software, Telecommunications and Computer Networks (SoftCOM)}, pages 1--3. IEEE, 2021.

\bibitem{hannousse2021securing}
Abdelhakim Hannousse and Salima Yahiouche.
\newblock Securing microservices and microservice architectures: A systematic mapping study.
\newblock {\em Computer Science Review}, 41:100415, 2021.

\bibitem{sultan2019container}
Sari Sultan, Imtiaz Ahmad, and Tassos Dimitriou.
\newblock Container security: Issues, challenges, and the road ahead.
\newblock {\em IEEE access}, 7:52976--52996, 2019.

\bibitem{ibrahim2019attack}
Amjad Ibrahim, Stevica Bozhinoski, and Alexander Pretschner.
\newblock Attack graph generation for microservice architecture.
\newblock In {\em Proceedings of the 34th ACM/SIGAPP symposium on applied computing}, pages 1235--1242, 2019.

\bibitem{yarygina2018overcoming}
Tetiana Yarygina and Anya~Helene Bagge.
\newblock Overcoming security challenges in microservice architectures.
\newblock In {\em 2018 IEEE Symposium on Service-Oriented System Engineering (SOSE)}, pages 11--20. IEEE, 2018.

\bibitem{rahman2023security}
Akond Rahman, Shazibul~Islam Shamim, Dibyendu~Brinto Bose, and Rahul Pandita.
\newblock Security misconfigurations in open source kubernetes manifests: An empirical study.
\newblock {\em ACM Transactions on Software Engineering and Methodology}, 32(4):1--36, 2023.

\bibitem{budigiri2021network}
Gerald Budigiri, Christoph Baumann, Jan~Tobias M{\"u}hlberg, Eddy Truyen, and Wouter Joosen.
\newblock Network policies in kubernetes: Performance evaluation and security analysis.
\newblock In {\em 2021 Joint European Conference on Networks and Communications \& 6G Summit (EuCNC/6G Summit)}, pages 407--412. IEEE, 2021.

\bibitem{nkomo2019software}
Peter Nkomo and Marijke Coetzee.
\newblock Software development activities for secure microservices.
\newblock In {\em Computational Science and Its Applications--ICCSA 2019: 19th International Conference, Saint Petersburg, Russia, July 1--4, 2019, Proceedings, Part V 19}, pages 573--585. Springer, 2019.

\bibitem{yu2019survey}
Dongjin Yu, Yike Jin, Yuqun Zhang, and Xi~Zheng.
\newblock A survey on security issues in services communication of microservices-enabled fog applications.
\newblock {\em Concurrency and Computation: Practice and Experience}, 31(22):e4436, 2019.

\bibitem{torkura2017integrating}
Kennedy~A Torkura, Muhammad~IH Sukmana, and Christoph Meinel.
\newblock Integrating continuous security assessments in microservices and cloud native applications.
\newblock In {\em Proceedings of the10th International Conference on Utility and Cloud Computing}, pages 171--180, 2017.

\bibitem{kubernetes2019kubernetes}
T~Kubernetes.
\newblock Kubernetes.
\newblock {\em Kubernetes. Retrieved May}, 24:2019, 2019.

\bibitem{microservices-demo}
Sock shop.
\newblock \url{https://github.com/microservices-demo/microservices-demo}, Last Access: Nov. 2023.

\bibitem{van2023continuous}
Andr{\'e} van Hoorn, Vincenzo Ferme, and Henning Schulz.
\newblock Continuous performance testing for microservices.
\newblock {\em HPI Future SOC Lab--Proceedings 2018}, (151):105, 2023.

\bibitem{nobre2023anomaly}
Jo{\~a}o Nobre, EJ~Solteiro Pires, and Ars{\'e}nio Reis.
\newblock Anomaly detection in microservice-based systems.
\newblock {\em Applied Sciences}, 13(13):7891, 2023.

\bibitem{inproceedings}
Muharrem Aksu, Mustafa~Hadi Dilek, Emin Tatlı, Kemal Bicakci, H.~Dirik, Mustafa Demirezen, and Tayfun Aykir.
\newblock A quantitative cvss-based cyber security risk assessment methodology for it systems.
\newblock pages 1--8, 10 2017.

\bibitem{wu2018weighted}
Zhenyu Wu, Yang Guo, Wenfang Lin, Shuyang Yu, and Yang Ji.
\newblock A weighted deep representation learning model for imbalanced fault diagnosis in cyber-physical systems.
\newblock {\em Sensors}, 18(4):1096, 2018.

\bibitem{aurelio2019learning}
Yuri~Sousa Aurelio, Gustavo~Matheus De~Almeida, Cristiano~Leite de~Castro, and Antonio~Padua Braga.
\newblock Learning from imbalanced data sets with weighted cross-entropy function.
\newblock {\em Neural processing letters}, 50:1937--1949, 2019.

\bibitem{tunstall2022natural}
Lewis Tunstall, Leandro Von~Werra, and Thomas Wolf.
\newblock {\em Natural language processing with transformers}.
\newblock " O'Reilly Media, Inc.", 2022.

\bibitem{devlin2018bert}
Jacob Devlin, Ming-Wei Chang, Kenton Lee, and Kristina Toutanova.
\newblock Bert: Pre-training of deep bidirectional transformers for language understanding.
\newblock {\em arXiv preprint arXiv:1810.04805}, 2018.

\bibitem{yang2019xlnet}
Zhilin Yang, Zihang Dai, Yiming Yang, Jaime Carbonell, Russ~R Salakhutdinov, and Quoc~V Le.
\newblock Xlnet: Generalized autoregressive pretraining for language understanding.
\newblock {\em Advances in neural information processing systems}, 32, 2019.

\bibitem{liu2019roberta}
Yinhan Liu, Myle Ott, Naman Goyal, Jingfei Du, Mandar Joshi, Danqi Chen, Omer Levy, Mike Lewis, Luke Zettlemoyer, and Veselin Stoyanov.
\newblock Roberta: A robustly optimized bert pretraining approach.
\newblock {\em arXiv preprint arXiv:1907.11692}, 2019.

\bibitem{Equations}
Karen~Scarfone S.~R. Peter~Mell.
\newblock A complete guide to the common vulnerability scoring system.
\newblock \url{https://www.first.org/cvss/v2/guide}, Last Access Dec 2023.

\bibitem{sparck1972statistical}
Karen Sparck~Jones.
\newblock A statistical interpretation of term specificity and its application in retrieval.
\newblock {\em Journal of documentation}, 28(1):11--21, 1972.

\bibitem{mikolov2013distributed}
Tomas Mikolov, Ilya Sutskever, Kai Chen, Greg~S Corrado, and Jeff Dean.
\newblock Distributed representations of words and phrases and their compositionality.
\newblock {\em Advances in neural information processing systems}, 26, 2013.

\bibitem{9513723}
Veneta Yosifova, Antoniya Tasheva, and Roumen Trifonov.
\newblock Predicting vulnerability type in common vulnerabilities and exposures (cve) database with machine learning classifiers.
\newblock In {\em 2021 12th National Conference with International Participation (ELECTRONICA)}, pages 1--6, 2021.

\bibitem{nvidia2023}
NVIDIA.
\newblock Train with mixed precision.
\newblock \url{https://docs.nvidia.com/deeplearning/performance/mixed-precision-training/index.html}, Last Access: Dec 2023.

\end{thebibliography}
\end{document}